\documentstyle[aps,epsf,rotate,multicol]{revtex}

\begin{document}

\draft

\title{Power Law Scaling for a System of Interacting Units with 
Complex Internal Structure}

\author{
Lu\'{\i}s~A.~Nunes Amaral$^{(1,2,3)}$, Sergey~V. Buldyrev$^{(2)}$,
Shlomo Havlin$^{(2,4)}$,\\ Michael A. Salinger$^{(5)}$, and H.~Eugene
Stanley$^{(2)}$ }

\address{
$^{(1)}$Theorie II, IFF, Forschungszentrum J\"ulich, 
 D-52425 J\"ulich, Germany \\
$^{(2)}$Center for Polymer Studies and Department of Physics,
 Boston University, Boston, MA 02215 \\
$^{(3)}$Condensed Matter Theory, Physics Dept., Massachusetts Institute
 of Technology, Cambridge, MA 02139 \\ 
$^{(4)}$Minerva Center and Department of Physics, Bar-Ilan University,
 Ramat Gan, Israel \\
$^{(5)}$Department of Finance and Economics, School of Management,
 Boston University, Boston, MA 02215
}

\date{Submitted: July 28, 1997; Printed: \today}

\maketitle

\begin{abstract}

 We study the dynamics of a system composed of interacting units each
with a complex internal structure comprising many subunits. We
consider the case in which each subunit grows in a multiplicative
manner. We propose a model for such systems in which the interaction
among the units is treated in a mean field approximation and the
interaction among subunits is nonlinear. To test the model, we
identify a large data base spanning 20 years, and find that the model
correctly predicts a variety of empirical results.

\end{abstract}


\begin{multicols}{2}

In the physical sciences, power law scaling is usually associated with
critical behavior, thus requiring a particular set of parameter
values.  For example, in the Ising model there is a particular value
of the strength of the interaction between the units composing the
system that generates correlations extending throughout the entire
system and leads to power law distributions \cite{SOC}. In the social
and biological sciences, there also appear examples of power law
distributions (incomes \cite{Pareto}, city sizes \cite{Zipf},
extinction of species \cite{Gould}, bird populations \cite{Keitt},
heart dynamics \cite{Peng}). However, it is difficult to imagine that
for all these diverse systems, the parameters controlling the dynamics
spontaneously self-tune to their critical values.

In this Letter, we raise an alternative mechanism by asking how power
law distributions can emerge even in the absence of critical dynamics.
The guiding principles for our approach, to be justified below, are:
(i) the units composing the system have a complex evolving structure
(e.g., the companies competing in an economy are composed of
divisions, the cities in a country competing for the mobile population
are composed of distinct neighborhoods, the population of some species
living in a given ecosystem might be composed of groups living in
different areas), and (ii) the size of the subunits composing each
unit evolve according to a random multiplicative process.

Fortunately, for one of the examples listed above, there is a wealth
of quantitative data, and here we focus on a large database giving the
time evolution of the size of companies \cite{Compustat}. In an
economy, the units composing the entire system are the competing
companies. In general, these companies have a complex internal
structure, with each company composed of divisions (the subunits of
each unit). It has been proposed that the evolution of a company's
size is described by a random multiplicative process with variance
independent of the size, and that each company can be viewed as a
structureless unit \cite{Gibrat}. However, later studies
\cite{Sutton,Gort,group70,group80,group90,Jovanovic,Stanley} reveal
that the dynamics of real companies are not fully consistent with the
simplified picture of Ref.~\cite{Gibrat}.

We develop here a model that dynamically builds a diversified,
multi-divisional structure, reproducing the fact that a typical
company passes through a series of changes in organization, growing
from a single-product, single-plant company, to a multi-divisional,
multi-product company \cite{Chandler}. The model reproduces a number
of empirical observations for a wide range of values of parameters and
provides a possible explanation for the robustness of the empirical
results. Due to our encouraging results for the case of company
growth, our model may offer a generic approach to explain power law
distributions in other complex systems.

The model, illustrated in Fig.~\ref{f-model}, is defined as follows.
A company is created with a single division, which has a size $\xi_1(t
= 0)$. The size of a company $S \equiv \Sigma_i \xi_i(t)$ at time $t$
is the sum of the sizes of the divisions $\xi_i(t)$ comprising the
company. We define a minimum size $S_{\rm min}$ below which a company
would not be economically viable, due to the competition between
companies; $S_{\rm min}$ is a characteristic of the industry in which
the company operates. We assume that the size of each division $i$ of
the company evolves according to a random multiplicative process
\cite{Gibrat}. We define
\begin{equation}
\Delta \xi_i(t) \equiv \xi_i(t)~\eta_{i}(t)\,,
\label{e-growth}
\end{equation}
where $\eta_{i}(t)$ is a Gaussian-distributed random variable with zero
mean and standard deviation $V$ independent of $\xi_i$. The divisions
evolve as follows:

\begin{itemize}
\item[(i)] If $\Delta \xi_i(t) < S_{\rm min}$, division $i$ evolves by
changing its size, and $\xi_i(t+1) = \xi_i(t) + \Delta \xi_i(t)$. If
its size becomes smaller than $S_{\rm min}$ --- i.e. if $\xi_i(t+1) <
S_{\rm min}$ --- then with probability $p_a$, division $i$ is
``absorbed'' by division $1$. Thus, the parameter $p_a$ reflects the
fact that when a division becomes very small it will 
\begin{figure}[htb]
\narrowtext
\centerline{
\epsfysize=0.65\columnwidth{\epsfbox{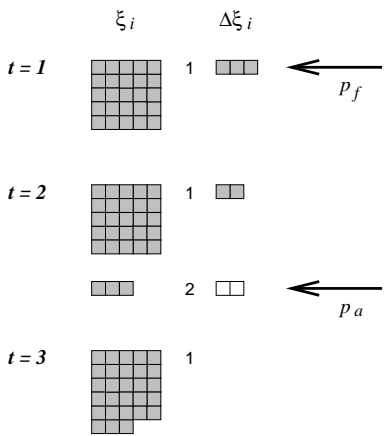}}}
\vspace*{0.5cm}
\caption{ Schematic representation of the time evolution of the size
and structure of a company. We choose $S_{\rm min} = 2$, and $p_f = p_a
= 1.0$. The first column of full squares represents the size $\xi_i$
of each division, and the second column represents the corresponding
change in size $\Delta \xi_i$. Empty squares represent negative
growth and full squares positive growth. We assume, for this example,
that the company has initially one division of size $\xi_1 = 25$,
represented by a $5 \times 5$ square. At $t=1$, division 1 grows by
$\Delta \xi_1 = 3$. A new division, numbered 2, is created because
$\Delta \xi_1 > S_{\rm min} = 2$, and the size of division 1 remains
unchanged, so for $t=2$, the company has 2 divisions with sizes $\xi_1 =
25$ and $\xi_2 = 3$. Next, divisions $\xi_1$ and $\xi_2$ grow by $2$
and $-2$, respectively. Division $2$ is absorbed by division $1$,
since otherwise its size would become $\xi_2 = 3 - 2 = 1$ which is
smaller than $S_{\rm min}$. Thus, at time $t=3$, the company has only
one division with size $\xi_1 = 25 + 2 + 1 = 28$. Note that if
division $1$ would be absorbed, then division $2$ would absorb
division $1$ and would then be renumbered $1$. If, division $1$ is
absorbed and there are no more divisions left, the company ``dies.'' }
\label{f-model}
\end{figure}
no longer be
viable due to the competition between companies.

\item[(ii)] If $\Delta \xi_i(t) > S_{\rm min}$, then with probability
$(1-p_f)$, we set $\xi_i(t+1) = \xi_i(t) + \Delta \xi_i(t)$. With a
probability $p_f$, division $i$ does not change its size --- so that
$\xi_i(t+1) = \xi_i(t)$ --- and an altogether new division $j$ is
created with size $\xi_j(t+1) = \Delta \xi_i(t)$. Thus, the parameter
$p_f$ reflects the tendency to diversify: the larger is $p_f$, the
more likely it is that new divisions are created.
\end{itemize}

\hspace*{-0.5cm} The dynamics are thus controlled by four parameters:
$S_{\rm min}$, $V$, $p_a$, and $p_f$; $S_{\rm min}$ just sets the
scale, so the results of the model do not depend on its value. We
assume that there is a broad distribution of values of $S_{\rm min}$
in the system because companies in different activities will have
different constraints.

\begin{figure}[htb]
\narrowtext
\centerline{
\epsfysize=0.9\columnwidth{\rotate[r]{\epsfbox{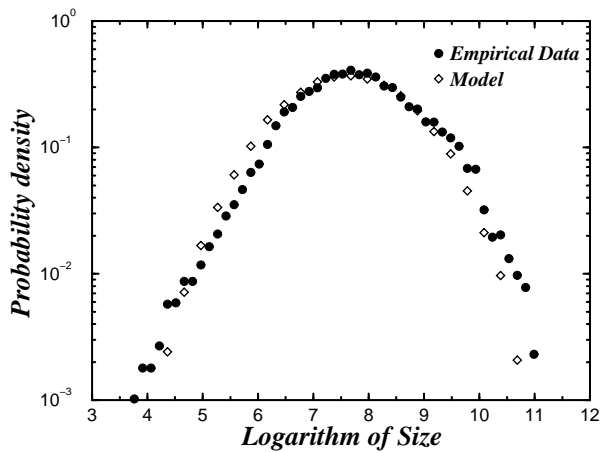}}}}
\vspace*{0.7cm}
\caption{ Probability density of the logarithm of company
size for the model and for US publicly-traded manufacturing firms in
the 1994 ``Compustat'' data base. To obtain these results, we assume
that $S_{\rm min}$ is drawn from a truncated log-normal distribution
with average value $5\times10^5$. The numerical simulations were
performed with parameters $V = 0.15$, $p_f = 0.8$, $p_a = 0.05$, and
$\ell = 50$ (for these parameter values, the actual probability of a
new division being created per division and per iteration is
approximately $0.01$). }
\label{f-dist-s}
\end{figure}

In Fig.~\ref{f-dist-s}, we compare the predictions of the model for
the distribution of company sizes with the empirical data
\cite{Stanley}.  We find similar results for a wide range of
parameters: $V = 0.1 - 0.2$, $p_a = 0.01 - 1$, and $p_f = 0.1 -
1.0$. We define 
one ``year'' to be $\ell$ iterations of our rules
applied to each company, and we find no significant dependence of the
results on the value of $\ell$ for $\ell = 20, 30$ or $50$.

\begin{figure}[tbh]
\narrowtext
\centerline{
\epsfysize=0.7\columnwidth{\rotate[r]{\epsfbox{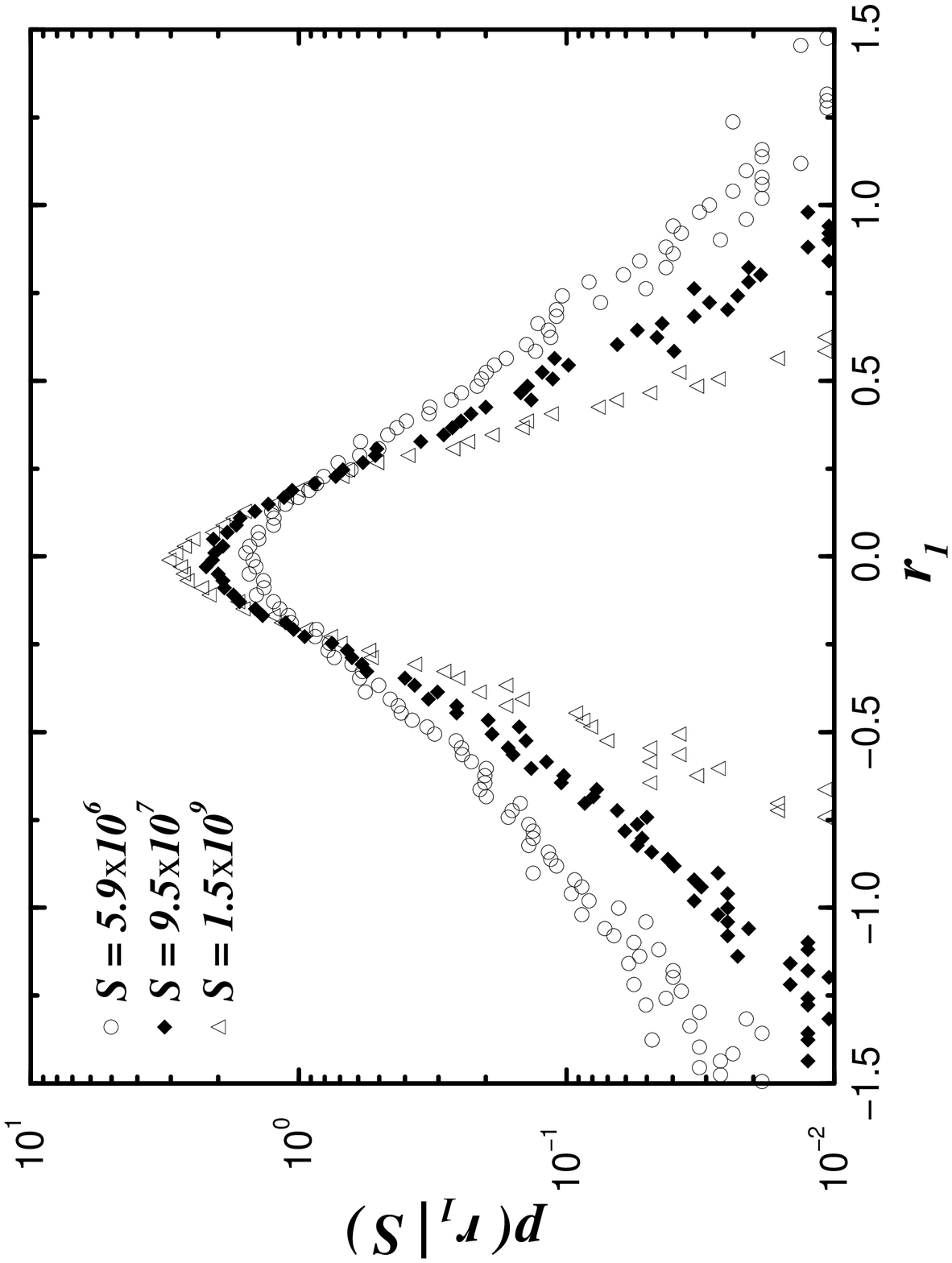}}}}
\vspace*{0.2cm}
\centerline{
\epsfysize=0.7\columnwidth{\rotate[r]{\epsfbox{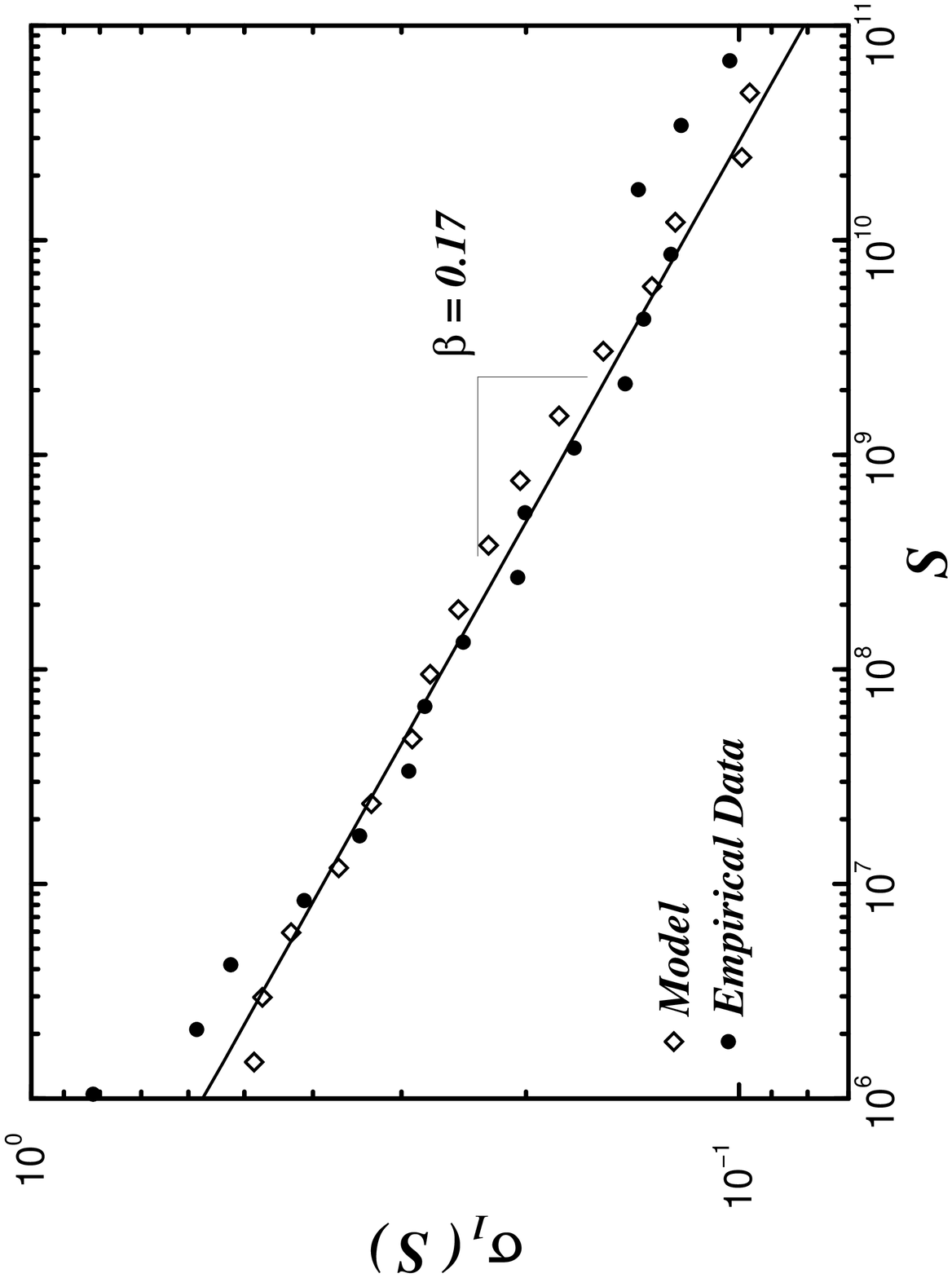}}}}
\vspace*{0.2cm}
\centerline{
\epsfysize=0.7\columnwidth{\rotate[r]{\epsfbox{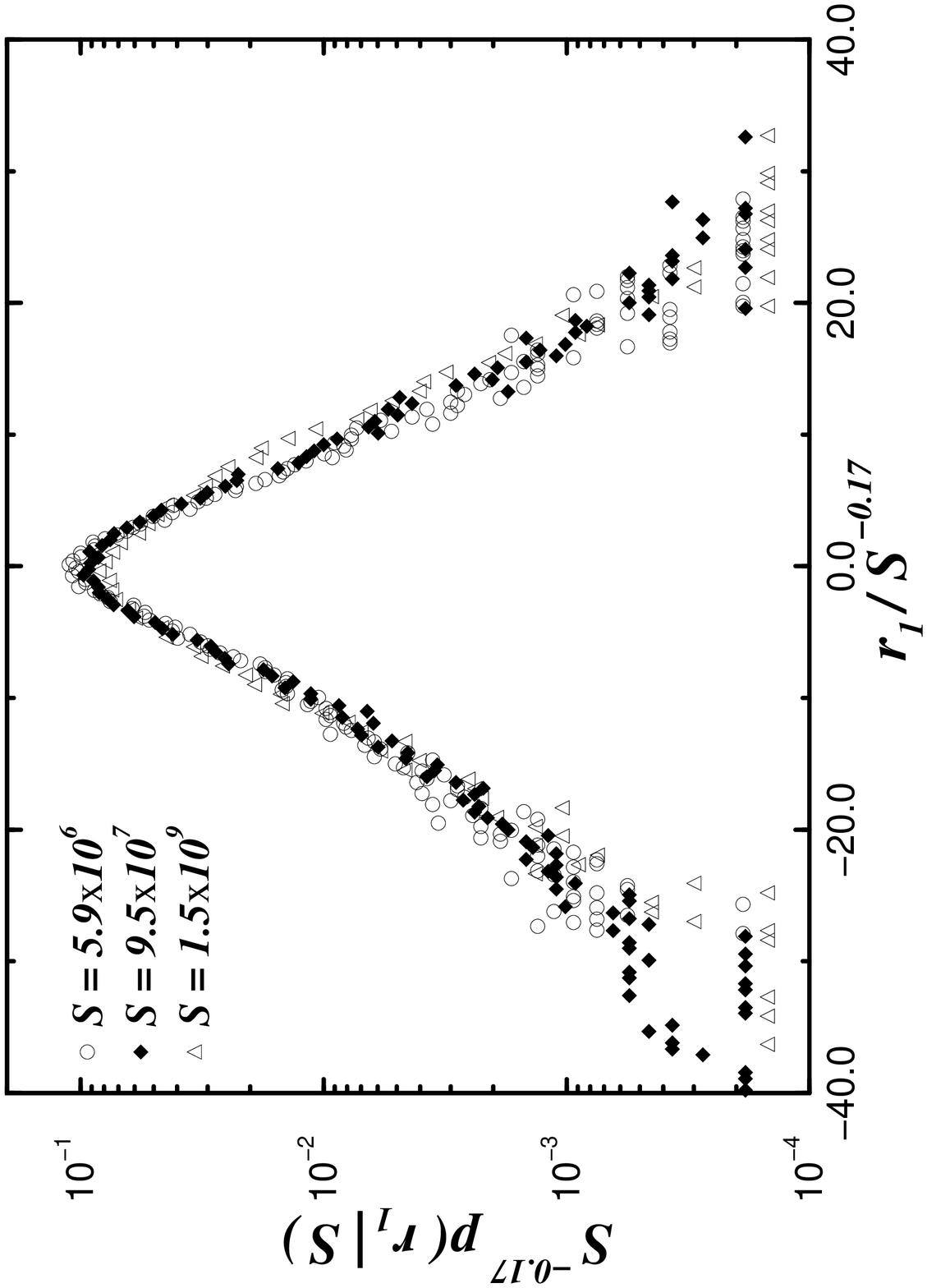}}}}
\vspace*{0.7cm}
\caption{ (a) Probability density of one-year growth rates
for firms with different sizes. The values of the parameters are the
same as in Fig.~\protect\ref{f-dist-s}. The distributions are
tent-shaped, as for the empirical data \protect\cite{Stanley},
consistent with an exponential distribution. (b) Dependence of the
standard deviation of the growth rates on company size. Shown are the
predictions of the model and the empirical results. The straight line
with slope $0.17$ is a least square fit to the predictions of the
model. (c) Probability density of one-year growth rates for different
company sizes plotted in scaled variables. }
\label{f-grates}
\end{figure}

It is common to study the logarithm of the one-year growth rate, $r_1
\equiv \ln R_1$, where $R_1 \equiv S(y+1) / S(y)$, and $S(y)$ and
$S(y+1)$ are the sizes of the company in the years $y$ and $y+1$. The
empirical distribution of $r_1$ for companies with size $S$ is
consistent with an exponential form \cite{Stanley}
\begin{equation}
p(r_1|S) = \frac{1}{\sqrt{2}\sigma_1(S)}
\exp\left(-{\sqrt{2}\,|r_1 - \bar r_1|\over\sigma_1(S)}\right)\,,
\label{e-distribution}
\end{equation}
where $\bar r_1$ represents the average growth rate. Moreover, the
standard deviation $\sigma_1(S)$ is consistent with a power law form
\begin{equation}
\sigma_1(S) \sim S^{-\beta},
\label{e-sigma}
\end{equation}
and for US manufacturing companies, $\beta \approx 0.2$
\cite{Stanley}. Figure~\ref{f-grates}a displays $p(r_1|S)$, and is
quite similar in form to empirical results \cite{Stanley}.
Figure~\ref{f-grates}b compares $\sigma_1(S)$ with the empirical data
of Ref.~\cite{Stanley}: for both, Eq.~(\ref{e-sigma}) holds with
$\beta = 0.17 \pm 0.03$.
Equations~(\ref{e-distribution})--(\ref{e-sigma}) allow us to scale
the growth rate distributions for different company sizes
(Fig.~\ref{f-grates}c).

\begin{figure}
\narrowtext
\centerline{
\epsfysize=0.7\columnwidth{\rotate[r]{\epsfbox{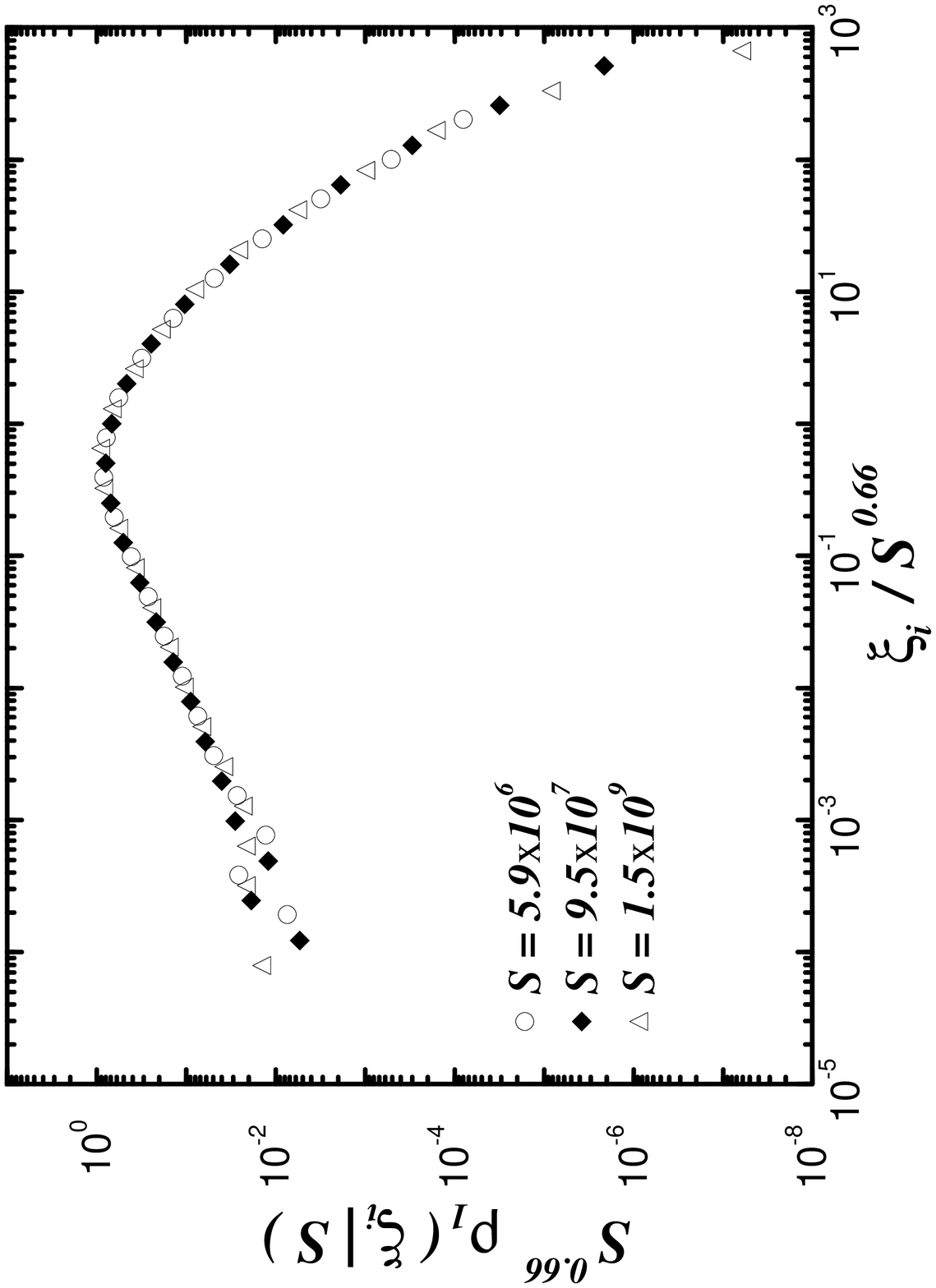}}}}
\vspace*{0.2cm}
\centerline{
\epsfysize=0.7\columnwidth{\rotate[r]{\epsfbox{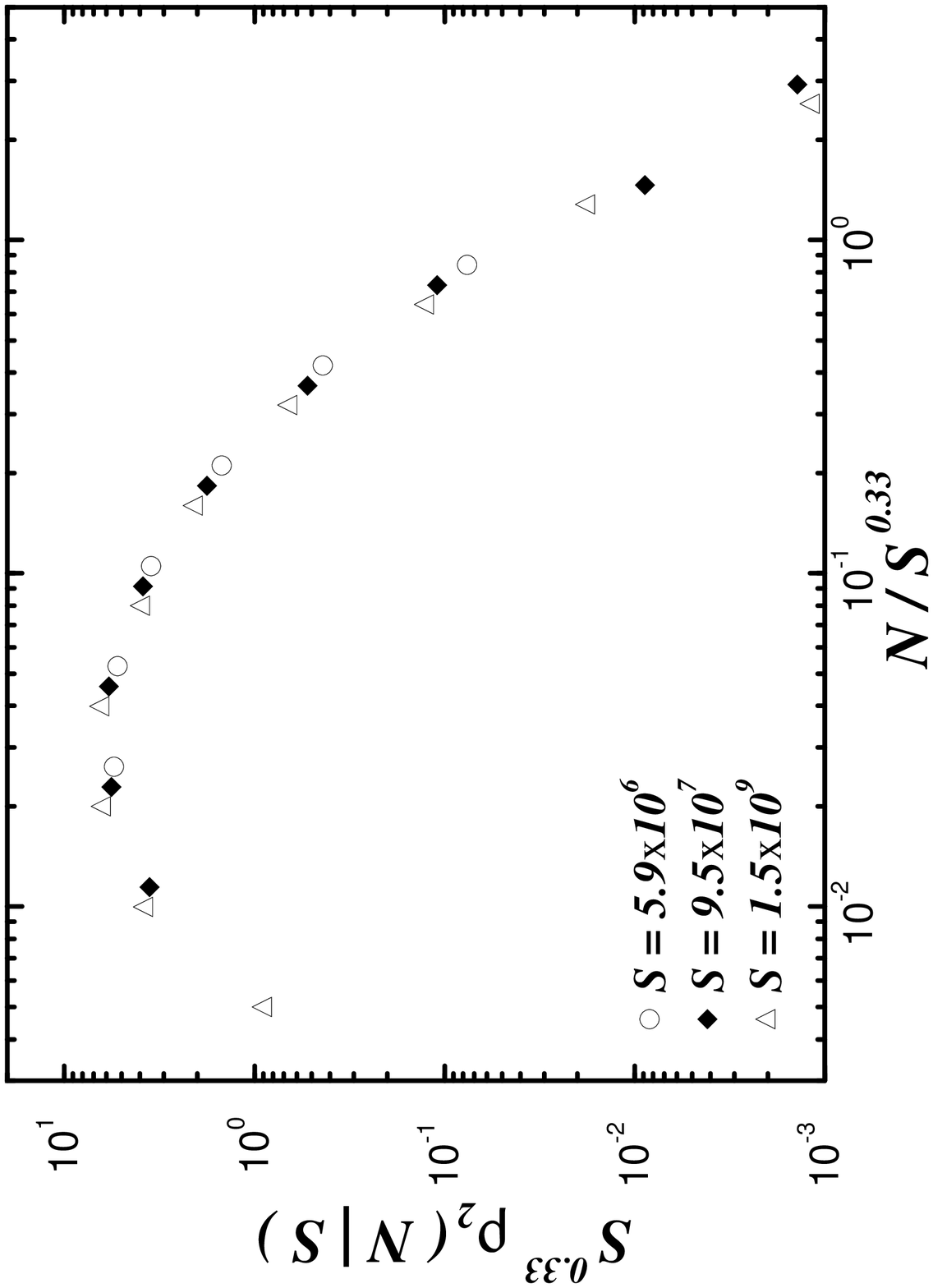}}}}
\vspace*{0.7cm}
\caption{ (a) ``Data collapse'' of the conditional probability density
$\rho_1$; the data fall onto a single curve corresponding to the scaling
form (\protect\ref{e-self}). (b) ``Data collapse'' of the conditional
probability density $\rho_2$; the data fall onto a single curve
corresponding to the scaling form (\protect\ref{e-self1}). }
\label{f-self-similar}
\end{figure}

 We next address the question of the structure of a given company.  To
this end, we calculate the probability density $\rho_1(\xi_i|S)$ to
find a division of size $\xi_i$ in a company of size $S$. For the
model, we find that the distribution $\rho_1$ is peaked at a maximum
which scales as $S^{\alpha}$. Hence, we make the hypothesis that
$\rho_1$ obeys the scaling relation
\begin{equation}
\rho_1(\xi_i|S) \sim S^{-\alpha} f_1 \left(\xi_i / S^{\alpha}
\right) \,.
\label{e-self}
\end{equation}
This hypothesis is confirmed by the scaling plot of
Fig.~\ref{f-self-similar}a. We find $\alpha = 0.66\pm0.05$ from
plotting the average value of $\xi_i$ against $S$. The same value of
$\alpha$ leads to the best scaling plot.

 Next, we make the hypothesis that the probability density
$\rho_2(N|S)$ to find a company with size $S$ composed of $N$
divisions, obeys the scaling relation
\begin{equation}
\rho_2(N|S) \sim S^{-(1-\alpha)} f_2 \left(N / S^{1-\alpha}
\right) \,.
\label{e-self1}
\end{equation}
In writing (\ref{e-self1}), we use the fact that from (\ref{e-self})
the characteristic size of a typical division scales as $S^{\alpha}$,
so that the typical number of divisions in a company is $S / S^{\alpha}
\sim S^{1-\alpha}$. Figure~\ref{f-self-similar}b shows that the
results of the model are consistent with the scaling relation
(\ref{e-self1}), with the same value of the scaling exponent $\alpha$
used in Fig.~\ref{f-self-similar}a.

The results described by Eqs.~(\ref{e-self})-(\ref{e-self1}) are in
qualitative agreement with empirical studies \cite{Gort,Jovanovic}
that show larger firms to be more diversified. Moreover,
Eqs.~(\ref{e-self})-(\ref{e-self1}) lead to the simple scaling law
\begin{equation}
\beta = (1 - \alpha) / 2 \,,
\label{e-beta}
\end{equation}
which can be tested. For $\alpha = 0.66\pm0.05$, (\ref{e-beta})
predicts $\beta = 0.17 \pm 0.03$, in remarkable agreement with our
independent calculation of $\beta$.

 We find that the predictions of the model are only weakly sensitive
to the parameter values, which perhaps is the reason why firms
operating in quite different industries are described by very similar
empirical laws. Accordingly, we conjecture that the scaling laws
found for US manufacturing firms \cite{Stanley} also hold for other
countries, such as Japan, with $\beta \approx 0.2$; this conjecture is
currently being tested with empirical data \cite{Takayasu}.

 The present model rests on a small number of assumptions. The two key
assumptions are: (i) Firms tend to organize themselves into multiple
divisions once they achieve a certain size. This assumption holds for
many modern corporations \cite{Chandler}. (ii) Growth rates of
different divisions are independent of one another. For an economist,
the latter is perhaps the stronger of the two assumptions. We find that
correlations in the growth rates of divisions within a same company,
even weak correlations, lead to $\beta \approx 0$. Thus, we confirm
that it is the assumption of independence among the growth rates that
reproduces the empirical findings of Refs.~\cite{Stanley}.

 There are two features of our results that are perhaps surprising.
First, although firms in our model consist of independent divisions, we
do not find $\beta = 1/2$. To understand why $\beta < 1/2$, suppose
that the distribution of $s_m \equiv \ln S_{\rm min}$ is a Dirac
$\delta$-function. Although this assumption is unrealistic, it leads to
an understanding of the underlying mechanisms in the model. For this
case, we find (i) that the distribution of company sizes is still close
to log-normal, with a width $\cal W$ which is a function of the
parameters of the model and, (ii) that the number of divisions increases
linearly with size, so $\alpha = 0$ and $\beta = 1/2$. Then, by
integration over $s_m$, we can estimate the value of $\beta$ for the
case of a broader distribution of $s_m$. Suppose that $s_m$ follows
some arbitrary distribution with width $\cal D$. Averaging
$\sigma_1^2(S)$ over this distribution, we find $\beta = {\cal W} / 2
({\cal D} + {\cal W})$. For a wide range of the values of the model's
parameters, ${\cal D} > {\cal W}$, and we find that $\beta$ is
remarkably close to the empirical value $\beta \approx 0.2$.

 Second, the distribution $p(r_1|S)$ is not Gaussian but ``tent''
shaped. We find this result arises from the integration of
nearly-Gaussian distributions of the growth rates over the
distribution of $S_{\rm min}$. For large values of $|r_1|$, the
saddle point approximation gives $p(r_1|S) \sim \exp(-\log^2 |r_1|)$,
which decays slower than exponentially, in qualitative agreement with
the model's predictions and with empirical observations. For $|r_1|
\ll 1$, $p(r_1|S)$ is approximately Gaussian, while for intermediate
values of $|r_1|$, the distribution decays exponentially. Our analytical
predictions are in agreement with the model and with empirical results.

 The model leads to a number of conclusions. First, it suggests
the deviations in the empirical data from predictions of the random
multiplicative process may be explained (i) by the diversification of
firms, i.e., firms are made up of interacting subunits; and (ii) by
the fact that different industries have different underlying scales,
i.e., there is a broad distribution of minimum scales for the survival
of a unit (for example, a car manufacturer must be much larger than a
software company).

 Second, the model suggests a possible explanation for the common
occurrence of power law distributions in complex systems. Our results
suggest that the empirically observed power law scaling does not
require some ``critical state'' of the system, but rather can arise
from a interplay between random multiplicative growth and the complex
structure of the units composing the system. Here we addressed the
case in which the interactions between the units can be treated in a
mean field way through the imposition of a minimum size for the
subunits. We believe that more complex interactions will still lead to
power law scaling, and that our model may offer a possible framework
for the the study of complex systems.

\end{multicols}


\begin{thebibliography}{99}
\bibitem{Compustat}
Specifically, for each company in the ``Compustat'' database of all US
publicly-traded firms, it is listed a variety of economic variables
such as sales, number of employees, assets, etc. In this work, we use
the generic term ``size'' to mean any of these quantities because a
precious study \protect\cite{Stanley} showed them to scale in an
identical way.
\end{thebibliography}
\end{document}